\documentclass[twocolumn]{aastex63}

\newcommand{\uat}[2]{\href{http://astrothesaurus.org/uat/#2}{#1 (#2)}}

\usepackage{color}
\usepackage{multirow}
\usepackage{amsmath}
\usepackage{textcomp}
\usepackage{booktabs}

\accepted{October 5, 2020}
\submitjournal{ApJ}

\shorttitle{SBF Calibration}
\shortauthors{Kim \& Lee}

\begin{document}

\title{ 
Calibration of Surface Brightness Fluctuations for Dwarf Galaxies
in the Hyper Suprime-Cam $gi$ Filter System
}

\correspondingauthor{Myung Gyoon Lee}
\email{mglee@astro.snu.ac.kr, yoojkim@astro.snu.ac.kr}

\author[0000-0003-1392-0845]{Yoo Jung Kim} 
\author[0000-0003-2713-6744]{Myung Gyoon Lee} 
\affiliation{Astronomy Program, Department of Physics and Astronomy, SNUARC,
Seoul National University, 1 Gwanak-ro, Gwanak-gu, Seoul 08826, Republic of Korea}

\begin{abstract}

{Surface brightness fluctuation (SBF) magnitudes are a powerful standard candle to measure distances to semi-resolved galaxies in the local universe,  a majority of which are dwarf galaxies that have often bluer colors than bright early-type galaxies.
We present an empirical $i-$band 
SBF calibration in a blue regime, $0.2 \lesssim (g-i)_0 \lesssim 0.8$ in the Hyper Suprime-Cam (HSC) magnitude system. We measure SBF 
magnitudes for 12 
nearby dwarf galaxies of various morphological types with archival HSC imaging data, and use their tip of the red giant branch (TRGB) distances to derive fluctuation - color relations. In order to subtract contributions of fluctuations 
due to young stellar populations, we use five different $g-$band magnitude masking thresholds, 
$M_{g,{\rm thres}} = -3.5, -4.0, -4.5, -5.0,$ and $-5.5$ mag. We find that the rms scatter of the linear fit to the relation is the smallest (rms = 0.16 mag) in the case of $M_{g,{\rm thres}} = -4.0$ mag, $\overline{M_i} = (-2.65\pm0.13)+ (1.28\pm0.24) \times (g-i)_0$. This scatter is much smaller than those in the previous studies (rms=0.26 mag), and is closer to the value for bright red galaxies (rms=0.12 mag).
This calibration is consistent with predictions from metal-poor simple stellar population models.
}

\end{abstract}

\section{Introduction}\label{sec_introduction}

With recent deep and wide ground-based imaging surveys, a large number of dwarf galaxies in the Local Volume (LV) have been newly discovered \citep[e.g.][]{carlin16, coh18, car20a, dav21}.
These dwarf galaxies have been used for various purposes, such as testing structure formation theories in small scales (e.g. "Missing Satellites" problem; \citet{kly99, moo99, bul17, mul20, car21}) and studying the environmental effects on dwarf galaxies \citep{carlin21, dav21}.
For these purposes, it is crucial to obtain precise distances to dwarf galaxies.

If the old stellar populations of a galaxy are resolved, one can obtain its distance 
from photometry of red giant branch stars \citep[the tip of the red giant branch (TRGB) method;][]{lee93}.
However, for farther galaxies, of which stellar populations are semi-resolved, the surface brightness flucutation (SBF) method can be useful \citep{ton88}.

Basically, the SBF technique measures spatial fluctuations in a galaxy image, resulting from Poisson statistics in the number of stars per resolution element \citep{ton88}. 
The fluctuation amplitudes due to old red giant branch (RGB) stars and asymptotic red giant branch (AGB) stars in a galaxy may be used as a standard candle. The technique has been mostly applied to massive early-type galaxies which consist mainly of 
old stellar populations. 

The fluctuation amplitudes are measured in Fourier space and  contributions from non-stellar contaminating sources such as background objects and globular clusters 
are subtracted. 
The fluctuation amplitudes are then converted to distances using empirical \emph{SBF calibrations}, which are relations between the fluctuation magnitude and integrated galaxy color \citep{ton90, jen03, bla09, jen15}. 

The color dependence of the fluctuation reflects
that the fluctuation magnitudes depend on stellar populations, as studied by stellar population models 
\citep[e.g.][]{liu00, bla01, jen03}. 
In the regime of massive galaxies with old stellar populations, redder galaxies which consist of more metal-rich stellar populations show fainter SBF magnitudes than bluer galaxies with lower metallicity. 
The relation between the integrated color and the fluctuation magnitude is
widely studied 
using the Hubble Space Telescope (HST) \citep{bla09, bla10, jen15} 
and ground-based telescopes \citep{can18, car19}.
The relation 
is known to be tight in the red color regime. 
For example, even with the ground-based 
Canada-France-Hawaii Telescope (CFHT), the rms scatter in the $i$-band is as small as 0.12 mag \citep{can18}.
The rms scatter is even smaller than 0.1 mag in the F814W band based on the HST data \citep{bla10}.

However, the relation in the bluer 
regime for dwarf galaxies is not a simple extrapolation of the redder counterpart. Several groups have explored the use of the SBF method to determine distances to dwarf elliptical galaxies, which are metal-poor and thus bluer, in the LV.
They found that the color dependence is less than those for bright red galaxies \citep{jer98, jer00, jer01,  can18} and the intrinsic scatter is larger \citep{mie06,bla09,jen15}, complicating an accurate distance measurement for dwarf galaxies.
The scatter is attributed by a larger sampling scatter \citep{bla09}, presence of younger stellar populations \citep{jer98}, or both \citep{gre21}.

In particular, larger uncertainties are introduced in measuring SBF of galaxies with young stellar populations \citep{jen15, car19}.
Young stars 
in star forming regions are often clumped and very luminous, introducing a significant fluctuation. 
In addition, dust and nebular emssion affect SBF signals.
Therefore, 
one should avoid using regions with young stellar populations in measuring the SBF.

Recently, \citet{car19} derived a new relation between $i-$band SBF magnitude and integrated $g-i$ color in the blue regime ($0.3 \lesssim (g-i)_0 \lesssim 0.8$) using 32 LV dwarf galaxies with TRGB distances measured with CFHT (28 galaxies) 
and HST (6 galaxies in \citet{coh18}).
With the inclusion of star-forming galaxies to the calibration sample, they extended the use of SBF to the regime of dwarf galaxies with young stars.
In measuring the SBF, they masked point sources brighter than a threshold chosen in the range $-6 < M_g, M_i <-4$ mag and removed contributions of fainter objects by subtracting fluctuation powers measured in nearby background fields. 
Using this relation between SBF magnitude and integrated color, they confirmed group memberships of multiple blue dwarf galaxies in the LV 
\citep{car19a, car21}. However, the rms scatter of their calibration is as large as 0.26 mag,  
leading to a 15\% uncertainty in distance measurements.

In this study, we present a new SBF calibration in 
the blue regime ($0.2 \lesssim (g-i)_0 \lesssim 0.8$) using dwarf galaxies with TRGB distances in archival Subaru/Hyper Suprime-Cam (HSC) imaging data. In particular, using a fixed masking threshold magnitude for contaminating sources, we reduced the rms scatter down to 0.16 mag. The paper is structured as follows. In Section \ref{sec_data}, we describe the HSC data and our selection of galaxies used for calibration. In Section \ref{sec_sbf}, we illustrate the methods of SBF distance measurements, with  details on masking contaminating sources. In Section \ref{sec_calibration}, we derive an $i$-band SBF calibration. 
In Section \ref{sec_discussion}, we compare our calibration with recent ground-based empirical calibrations and with stellar population models. In Section \ref{sec_summary}, we summarize our primary results. 


\section{Data and Galaxy Sample}\label{sec_data}


For the SBF calibration, we 
search for dwarf galaxies that have HSC images with sufficient $g-$ and $i-$band exposures in the SMOKA Science Archive\footnote{https://smoka.nao.ac.jp/}.
Among them we selected nearby galaxies with TRGB distances smaller than 10 Mpc found in the Updated Nearby Galaxy Catalog \citep{kar13b}.
Galaxies with extremely low surface brightness (central surface brightness fainter than 24.5 mag arcsec$^{-2}$) were excluded from the sample. Moreover, galaxies of which the area bright enough to measure SBF is smaller than 50 arcsec$^{2}$ were excluded as well. 
In total, we selected 12 dwarf galaxies. 

\begin{figure*} 
\centering
\includegraphics[scale=0.55]{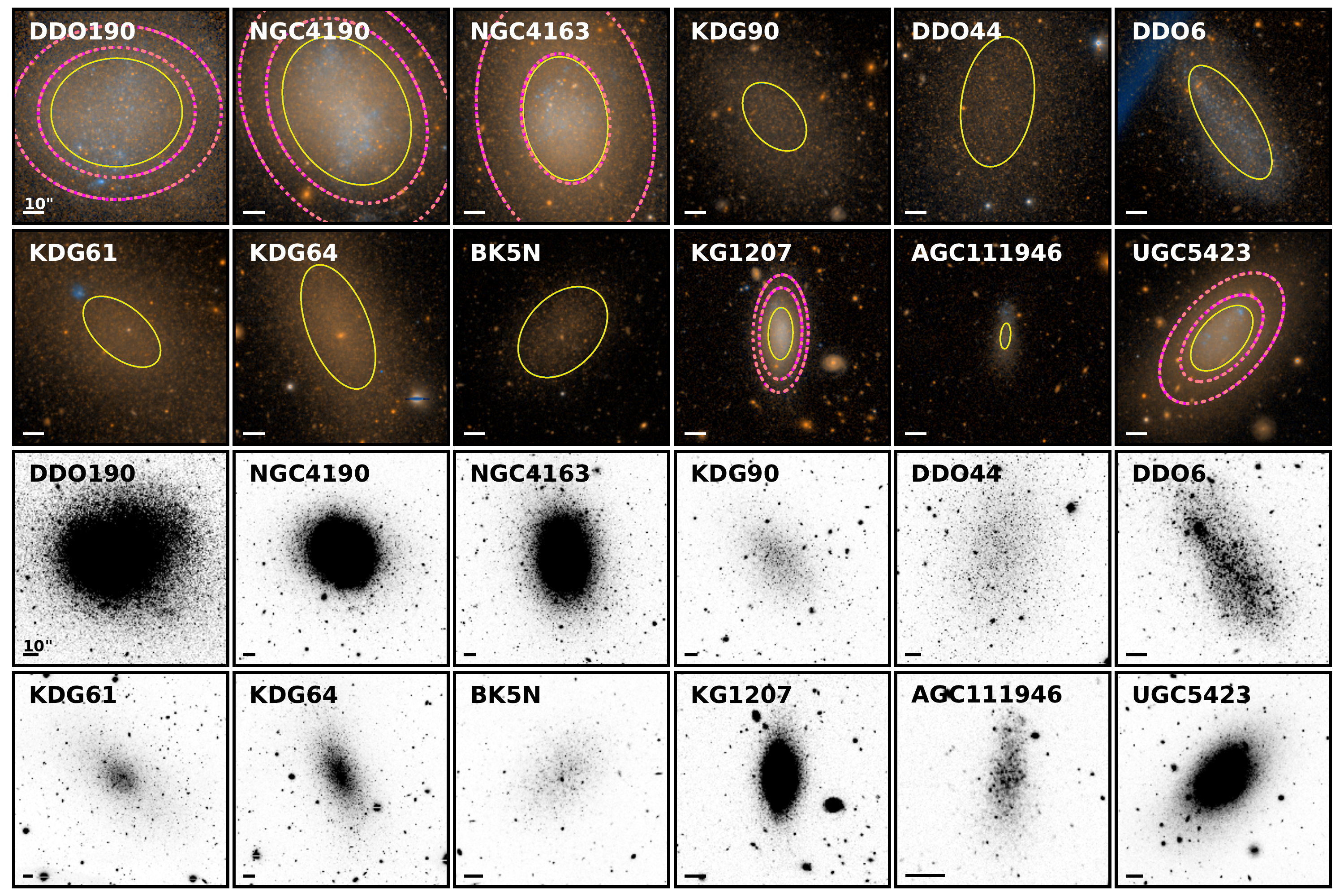} 
\caption{HSC images of our selected dwarf galaxies for the SBF calibration. (Top) $2\arcmin\times2\arcmin$  pseudo-color images based on $g-$ and $i-$band images. Solid lines indicate the boundary of the inner field in a galaxy and dotted  lines indicate that of the outer field. (Bottom) Gray scale $i-$band images of the same galaxies but with different fields of view to show better faint stellar light. North is up and east to the left. The scale bars in each image represent 10 arcsec.
}
\label{fig_sample}
\end{figure*}

\begin{deluxetable*}{lccccccc} 
\tablecaption{List of Dwarf Galaxies Used for SBF Calibration \label{tab_list}}
\tablehead{
\colhead{Name} & \colhead{R.A.} & \colhead{Decl.} & \colhead{D$_{TRGB}$\textsuperscript{a}} & \colhead{$M_B$\textsuperscript{b}} &  \colhead{$A_i$\textsuperscript{c}} & \colhead{Morphology\textsuperscript{d}} & \colhead{$t_{exp}$ ($g/i$)\textsuperscript{e}}\\
& (J2000) & (J2000) & (Mpc) & (mag) & (mag) & (mag) &  (s)
}
\startdata
DDO 190 & $14:24:43.5$ & $44:31:33$ & $2.76\pm0.08$ & $-14.1$ &  0.02 & Im & 810/1230 \\ 
NGC 4190 & $12:13:44.6$ & $36:38:00$ & $2.84\pm0.11$ & $-13.9$ &  0.05 & BCD & 7500/2580 \\ 
NGC 4163 & $12:12:08.9$ & $36:10:10$ & $2.95\pm0.08$ & $-13.8$ &  0.03 & Im & 14070/5430 \\ 
KDG 90 & $12:14:57.9$ & $36:13:08$ & $3.02\pm0.12$ & $-11.5$ &  0.03 & Tr & 7500/2580 \\ 
DDO 44 & $07:34:11.3$ & $66:53:10$ & $3.25\pm0.10$ & $-12.1$ &  0.07 & Sph & 3150/1350 \\ 
DDO 6 & $00:49:49.3$ & $-21:00:58$ & $3.40\pm0.15$ & $-12.4$ &  0.03 & Ir & 10935/2595 \\ 
KDG 61 & $09:57:02.7$ & $68:35:30$ & $3.63\pm0.10$ & $-12.9$ &  0.12 & Sph & 13336/13654 \\ 
KDG 64 & $10:07:01.9$ & $67:49:39$ & $3.71\pm0.10$ & $-12.6$ &  0.09 & Tr & 4430/6330 \\ 
BK5N & $10:04:40.3$ & $68:15:20$ & $3.72\pm0.17$ & $-10.6$ &  0.11 & Sph & 5530/7590 \\ 
KG 1207 & $12:09:56.4$ & $36:26:07$ & $4.88\pm0.32$ & $-13.0$ &  0.04 & Im & 5550/1710 \\ 
AGC 111946 & $01:46:41.6$ & $26:48:05$ & $8.36\pm0.29$ & $-11.4$ &  0.14 & Ir & 3150/2150 \\ 
UGC 5423 & $10:05:30.6$ & $70:21:52$ & $8.71\pm0.24$ & $-15.6$ &  0.14 & Ir & 4430/6330 \\ 
\enddata
\tablecomments{
\textsuperscript{a} TRGB distances calculated from F814W TRGB magnitudes in EDD \citep{jac09, ana21} with the recent blue-color TRGB calibration in \citet{jan21}. \\
\textsuperscript{b} $B-$band absolute total magnitudes from the Updated Nearby Galaxy Catalog \citep{kar13b}. \\
\textsuperscript{c} Foreground reddening based on the extinction maps by \citet{sch98} and the conversion coefficients obtained by \citet{sch11}. \\ 
\textsuperscript{d} Dwarf galaxy morphological types classified by \citet{kar13b} (Sph for spheroidal galaxies, Im for Magellanic type irregular galaxies, Ir for irregular galaxies, BCD for blue compact dwarfs, and Tr for transition types between spheroidals and irregulars.) 
\\
\textsuperscript{e} Exposure times for HSC $g$ and $i$-band images. 
}
\end{deluxetable*}


Figure \ref{fig_sample} shows pseudo color images based on HSC $g$ and $i$-band images (top panels) and gray scale $i-$band images (bottom panels) of the selected galaxies in our sample.
In Table \ref{tab_list} 
we list basic information of the galaxies sorted by the TRGB distances.
F814W TRGB magnitudes of all the galaxies were obtained from Extragalactic Distance Database (EDD)\citep{tul09, jac09, ana21}\footnote{http://edd.ifa.hawaii.edu/}. 
Since the dwarf galaxies have blue TRGB colors, we use the most recent color-independent TRGB calibration in \citet{jan21} (${M}_{{\rm{F}}814{\rm{W}}}^{{\rm{TRGB}}}=-4.050\pm 0.028$ (stat) $\pm 0.048$ (sys)) to convert TRGB magnitudes to distances. 
Note that the TRGB calibration we use in this study is consistent with the calibration in \citet{bla21} in tying the SBF method to the TRGB for massive early-type galaxies. 
The TRGB distances to these galaxies are from 2.76 to 8.71 Mpc.
Total magnitudes of the galaxies range from $M_B\approx -10.6$ to --15.6 mag.
Our sample is composed of various morphological types: 3 spheroidal galaxies, 7 late-type galaxies (BCD, Ir, and Im), and 2 transition types between early-type and late-type galaxies.

Among our sample, five galaxies (DDO 190, NGC 4190, NGC 4163, KG 1207, and UGC 5423) show a notable presence of young blue stars and higher surface brightness in the central regions compared with the outer regions. 
Thus, the stellar populations in the inner and the outer parts of these galaxies are likely to differ. We divide each of these galaxies into an inner field and an outer halo field. The areas of the chosen fields are indicated as solid lines and dotted lines in the top panels of Figure \ref{fig_sample}.
Therefore, the sample for SBF measurements consists of 17 fields in 12 galaxies. 

The HSC images are fully reduced, sky subtracted, and coadded with the pipeline \texttt{hscPipe} designed for LSST (\citet{bos18,ive19}). 
Global sky subtraction was carried out 
in two steps in \texttt{hscpipe v6} \citep{aih19}. Empirical background model was made by averaging pixels (ignoring detected objects) in 1024 $\times$ 1024 superpixel grids in the entire focal plane, then fitting the values with a 6th order two-dimensional Chebyshev polynomial. Smaller scale features were subtracted by fitting a scaled sky frame, which is the mean response of the instrument to the sky for a particular filter. This sky subtraction technique preserves low surface brightness features better than previous versions of \texttt{hscpipe}. 
The pixel scale of HSC images is 0.168{\arcsec} per pixel, and  
$i-$band seeing for these images are 0.5{\arcsec}-0.7{\arcsec}.

For the following analysis, we correct for foreground reddening on each galaxy based on the extinction maps by \citet{sch98} and the conversion coefficients obtained by \citet{sch11}, as listed in Table \ref{tab_list}.  All the magnitudes are in the AB system  and extinction- corrected magnitudes and colors are denoted with subscript 0 throughout the paper.


\section{SBF Measurement}\label{sec_sbf}

In this section, we describe our method 
of applying SBF techniques to the sample galaxies. We  present the SBF measurements 
for two example galaxies as shown in Figure \ref{fig_SBF}: one spheroidal galaxy (BK5N) and one irregular galaxy (DDO 6), both 
located at similar distances (3.40 Mpc and 3.72 Mpc, respectively.)
DDO 6 is the bluest and BK5N is the reddest in the sample.

\subsection{SBF Method}\label{sec_sbf_overall}

\begin{figure*} [h]
\centering
\includegraphics[scale=0.65]{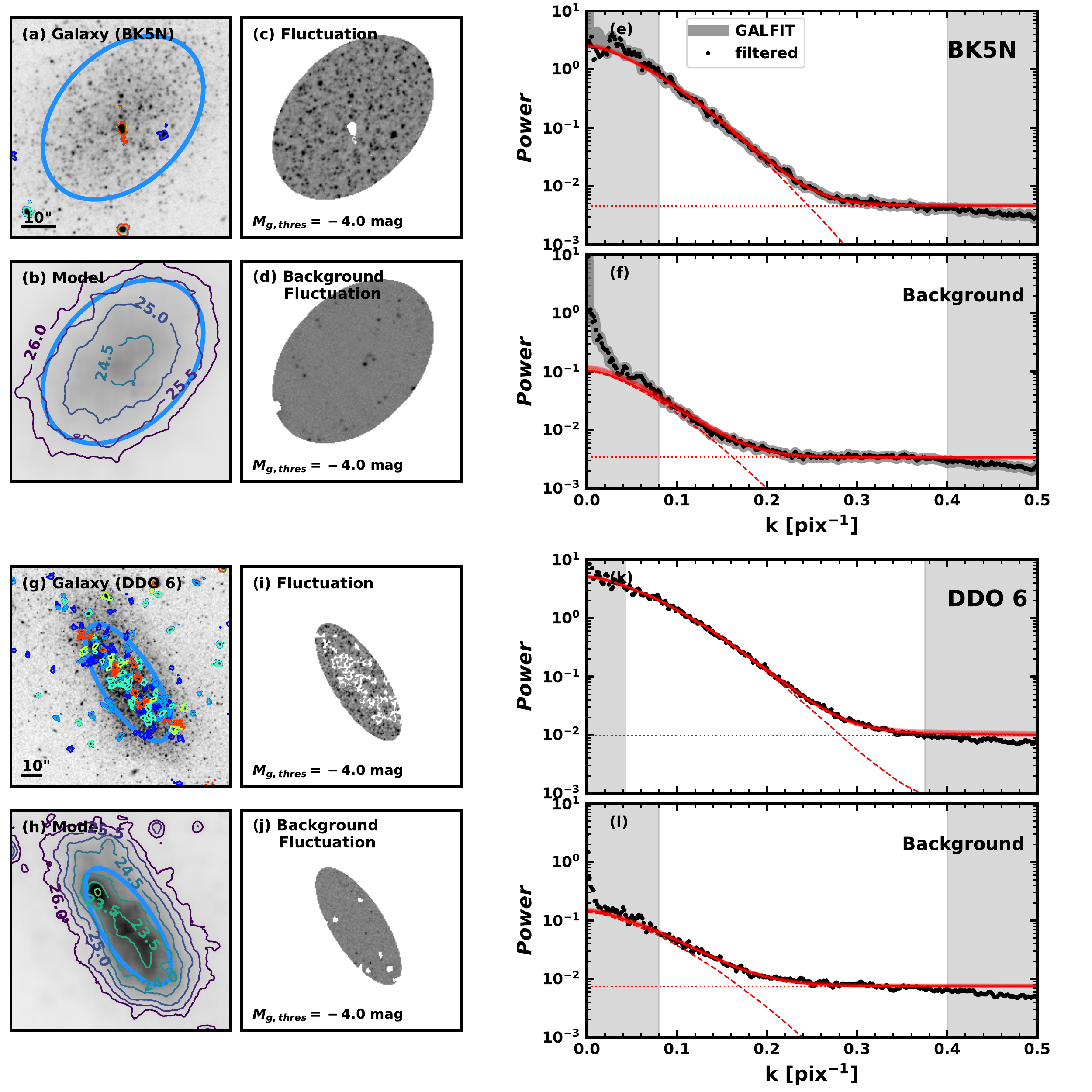} 
\caption{A summary of the components that go into the SBF measurement procedures 
for an early-type galaxy BK5N and an asymmetric late-type galaxy DDO 6. The upper and lower panels show the SBF measurement of BK5N and DDO 6, respectively. 
Panels (a) and (g) display $i$-band HSC images with blue ellipses representing the area used for SBF calculation. The  small contours (dark blue, light blue, cyan, light green, and orange) indicate sources masked using thresholds of $M_{g,{\rm thres}} = -3.5, -4.0, -4.5, -5.0,$ and $-5.5$ mag. 
Black bars represent 10{\arcsec} length. Panels (b) and (h) show median-filtered model images used as smooth galaxy models. Contours show the surface brightness as an unit of mag arcsec$^{-2}$. Panels (c) and (i) are fluctuation images, which are calculated as ${({\rm Galaxy} - {\rm Model})} / {\sqrt{\rm Model}}$. Panels (d) and (j) are examples of background fluctuations sampled from a random area with no apparent bright objects and calculated as $({\rm Background}) / {\sqrt{\rm Model}}$. Panels (e) and (k) show azimuthally averaged power spectra of galaxy fluctuation images (black circles for median-filtered model images, and gray thick lines for \texttt{GALFIT} galaxy model images) and fitted lines (red solid lines). 
The units of power are arbitrary.
Dashed lines and dotted lines indicate PSF-scale components and white noise components, respectively. Panels (f) and (l) show similar power spectra but for flat background fields within a radius of 0.2 degrees from the target galaxy. 
Gray regions at both ends are not used for fitting.}
\label{fig_SBF}
\end{figure*}

The SBF distance indicator was first introduced by \citet{ton88} and has been  widely used to measure distances to semi-resolved galaxies. The overall procedure of SBF measurements is well described in \citet{ton88}.
The fluctuation image is calculated as 
${({\rm Galaxy} - {\rm Model})} / {\sqrt{\rm Model}}$ (see Fig \ref{fig_SBF}(c) and (i) for BK5N and DDO 6, respectively), where Galaxy denotes the observed galaxy image (Fig \ref{fig_SBF}(a) and (g)) and Model corresponds to the smooth galaxy light component (Fig \ref{fig_SBF}(b) and (h)).
Since our sample is composed of not only early-type galaxies but also late-type galaxies,
we use median-filtered galaxy images as described below in Section \ref{sec_sbf_modelling} to obtain smooth galaxy light components instead of function-fitting galaxy models which have been often adopted for early-type galaxies (see Section \ref{sec_sbf_modelling}).

The one-dimensional azimuthally averaged power spectrum is obtained from the fluctuation image in the Fourier domain (see Fig \ref{fig_SBF}(e) and (k)).
The power spectrum $P(k)$ can be decomposed into 
the sum of a white noise component and the power spectrum of the point-spread function (PSF), scaled to match the stellar fluctuation amplitude.
Therefore, it is fit by a linear function, 
\begin{equation}\label{ps}
P(k) = P_{0} E(k) + P_{1}.
\end{equation}
E(k) is the expectation power spectrum, which is the convolution of the normalized PSF power spectrum and the power spectrum of the mask. The mask is explained in Section \ref{sec_sbf_mask} in detail. 
$P_{1}$ represents the white-noise component and $P_{0}$ represents the SBF amplitude, our quantity of interest. 
Right panels in Figure \ref{fig_SBF} show that the $P_{1}$ is not constant but slightly sloped at the large wavenumbers.
This is because pixel interpolations took place in stacking images by \texttt{hscpipe}. For this reason, fitting the largest wavenumbers should be avoided, as will be discussed in Section \ref{sec_sbf_fit}.

The apparent SBF magnitude is calculated by $\overline{m} = -2.5 \log{P_0} + zp$ where $zp$ is the photometric zeropoint.
The absolute SBF magnitude, $\overline{M_i}$, is obtained from fluctuation - integrated color relations, or SBF calibrations, derived empirically or theoretically \citep{ton90, bla10, jen15, can18}.
Thus, once SBF $P_0$ signals are measured and the integrated galaxy color is known, one can obtain galaxy distances from apparent SBF magnitudes. 

In measuring the fluctuation power, $P_0$, one needs to mask or subtract 
sources that can contaminate the SBF, namely globular clusters, foreground stars, and background galaxies.
The standard method which is applied mostly to early-type galaxies takes a statistical approach \citep{ton88, jen98, mei05, can05, can18}. 
First, find and apply masks to sources that are brighter than a certain magnitude threshold. Second, fit the luminosity functions of the sources with the combination of the globular cluster luminosity function (GCLF) and the background galaxy luminosity function. Third, extrapolate the luminosity functions to fainter magnitudes and calculate the contribution from unmasked sources. This power is then subtracted from the measured SBF power $P_0$ to derive a corrected apparent SBF magnitude. This method is reasonable for massive galaxies 
with sufficiently large number of globular clusters that the luminosity function can be determined reliably.

However, our sample galaxies are low-mass galaxies and many are late-types. The number of globular clusters in these galaxies is not large enough to derive a reasonable GCLF, requiring modification to the standard method. Recently, there have been efforts to obtain SBF distances to blue-regime dwarf galaxies which have young and blue stellar populations or low surface brightness \citep{van18, coh18, car19, car21}. 
Instead of modelling the contribution from globular clusters and background galaxies, \citet{car19}
adopted a different way utilizing the background field fluctuation. They mask sources brighter than a magnitude threshold and 
estimate the contribution of the remaining faint sources by measuring the fluctuations in nearby background fields. Then they subtract it from the measured galaxy fluctuation. Examples of background fluctuation images for BK5N and DDO 6 are shown in Figure \ref{fig_SBF}(d) and (j) and corresponding background power spectra in (f) and (l).

In this study, we follow the standard SBF measurement method \citep{ton88,mei05}, but conduct masking and background subtraction similar to the method used by \citet{car19}. 
For galaxy modelling, masking contaminating sources, and power spectra fitting, we provide a detailed description in the following subsections.

\begin{deluxetable*}{llcccccccc}[h]
\tabletypesize{\footnotesize}
\tablecaption{SBF Results for the Dwarf Galaxies \label{tab_calibration} }
\tablewidth{0pt}
\tablehead{
\colhead{Name} & & \colhead{D$_{TRGB}$}& \colhead{$(g - i)_0 ^{\dagger}$} & \colhead{$\overline{\mu_{i,0}} ^{\dagger}$} &
\multicolumn{5}{c}{SBF absolute magnitude $\overline{M_i}$ (mag)} \\ \cmidrule(lr){6-10} & & \colhead{(Mpc)} & \colhead{(mag)} & \colhead{(mag arcsec$^{-2}$)}& \multicolumn{5}{c}{Contaminating source masking threshold $M_{g,{\rm thres}}$(mag)} \\ &  &  & & &\colhead{--3.5} & \colhead{--4.0} & \colhead{--4.5} & \colhead{ --5.0} & \colhead{--5.5}
}
\startdata
\multirow{2}{*}{DDO 190} & inner  &  $2.76\pm0.08$  &  $0.29\pm0.03$  &  $22.25\pm0.02$  &  $-2.04\pm0.02$ & $-2.20\pm0.02$ & $-2.31\pm0.02$ & $-2.44\pm0.03$ & $-2.46\pm0.04$  \\ 
 & outer  &  $2.76\pm0.08$  &  $0.45\pm0.04$  &  $23.39\pm0.03$  &  $-1.94\pm0.02$ & $-1.97\pm0.02$ & $-1.93\pm0.06$ & $-2.59\pm0.04$ & $-2.62\pm0.04$  \\ 
\multirow{2}{*}{NGC 4190} & inner  &  $2.84\pm0.11$  &  $0.33\pm0.02$  &  $21.90\pm0.02$  &  $-2.39\pm0.04$ & $-2.40\pm0.04$ & $-2.42\pm0.04$ & $-2.48\pm0.04$ & $-2.56\pm0.04$  \\ 
 & outer  &  $2.84\pm0.11$  &  $0.58\pm0.03$  &  $24.25\pm0.02$  &  $-1.77\pm0.04$ & $-1.76\pm0.07$ & $-1.88\pm0.08$ & $-1.86\pm0.11$ & $-1.79\pm0.17$  \\ 
\multirow{2}{*}{NGC 4163} & inner  &  $2.95\pm0.08$  &  $0.47\pm0.02$  &  $21.92\pm0.02$  &  $-1.87\pm0.03$ & $-1.91\pm0.04$ & $-2.32\pm0.03$ & $-2.48\pm0.04$ & $-2.54\pm0.05$  \\ 
 & outer  &  $2.95\pm0.08$  &  $0.67\pm0.02$  &  $23.50\pm0.02$  &  $-1.78\pm0.03$ & $-1.78\pm0.05$ & $-2.30\pm0.03$ & $-2.42\pm0.05$ & $-2.42\pm0.06$  \\ 
KDG 90 &   &  $3.02\pm0.12$  &  $0.59\pm0.03$  &  $24.00\pm0.02$  &  $-2.21\pm0.03$ & $-2.20\pm0.03$ & $-2.19\pm0.04$ & $-2.19\pm0.05$ & $-2.17\pm0.06$  \\ 
DDO 44 &  &  $3.25\pm0.10$  &  $0.63\pm0.03$  &  $24.28\pm0.02$  &  $-2.02\pm0.08$ & $-1.98\pm0.17$ & $-1.94\pm0.32$ & $-2.03\pm0.30$ & $-2.00\pm0.35$  \\ 
DDO 6 &  &  $3.40\pm0.15$  &  $0.23\pm0.03$  &  $23.65\pm0.02$  &  $-2.16\pm0.04$ & $-2.42\pm0.04$ & $-2.60\pm0.04$ & $-2.86\pm0.05$ & $-3.15\pm0.06$  \\ 
KDG 61 &  &  $3.63\pm0.10$  &  $0.76\pm0.03$  &  $23.50\pm0.02$  &  $-1.50\pm0.04$ & $-1.49\pm0.04$ & $-1.49\pm0.04$ & $-1.49\pm0.04$ & $-1.48\pm0.05$  \\ 
BK5N &  &  $3.71\pm0.10$  &  $0.81\pm0.02$  &  $23.36\pm0.02$  &  $-1.73\pm0.03$ & $-1.73\pm0.03$ & $-1.73\pm0.03$ & $-1.72\pm0.04$ & $-1.71\pm0.04$  \\ 
KDG 64 &  &  $3.72\pm0.17$  &  $0.81\pm0.03$  &  $24.71\pm0.02$  &  $-1.71\pm0.06$ & $-1.72\pm0.09$ & $-1.70\pm0.10$ & $-1.69\pm0.12$ & $-1.69\pm0.12$  \\ 
\multirow{2}{*}{KG 1207} & inner  &  $4.88\pm0.32$  &  $0.33\pm0.03$  &  $21.57\pm0.02$  &  $-2.15\pm0.05$ & $-2.16\pm0.05$ & $-2.17\pm0.05$ & $-2.20\pm0.04$ & $-2.35\pm0.03$  \\ 
 & outer  &  $4.88\pm0.32$  &  $0.57\pm0.04$  &  $24.19\pm0.03$  &  $-1.52\pm0.05$ & $-1.67\pm0.07$ & $-1.65\pm0.07$ & $-1.64\pm0.08$ & $-1.80\pm0.10$  \\ 
\multirow{2}{*}{UGC 5423} & inner  &  $8.36\pm0.29$  &  $0.27\pm0.06$  &  $23.53\pm0.04$  &  $-2.46\pm0.05$ & $-2.46\pm0.06$ & $-2.59\pm0.04$ & $-2.79\pm0.05$ & $-2.79\pm0.03$  \\ 
 & outer  &  $8.71\pm0.24$  &  $0.42\pm0.02$  &  $22.24\pm0.02$  &  $-2.09\pm0.05$ & $-2.39\pm0.04$ & $-2.44\pm0.03$ & $-2.53\pm0.03$ & $-2.61\pm0.03$  \\ 
AGC 111946 &  &  $8.71\pm0.24$  &  $0.62\pm0.03$  &  $23.70\pm0.02$  &  $-1.76\pm0.07$ & $-1.77\pm0.07$ & $-1.80\pm0.08$ & $-1.80\pm0.10$ & $-1.80\pm0.11$  \\ 
\enddata
\tablecomments{$^{\dagger}$ Calculated in the area used for SBF calculation, with $M_{g,{\rm thres}}= -4.0$ mag. Errors include HSC calibration error 0.017 mag \citep{aih19} per each band. }
\end{deluxetable*}


\subsection{Galaxy Modelling}\label{sec_sbf_modelling}

As can be seen in Figure \ref{fig_sample}, many of our sample galaxies deviate from simple S\'ersic models due to star-forming regions and asymmetric features. For example, while BK5N (Figure \ref{fig_SBF}(a)) is well described by a single S\'ersic model, DDO 6 (Figure \ref{fig_SBF}(g)) has more complicated profiles. Therefore, instead of using simple galaxy models, we use median-filtered images as smooth galaxy models (Figure \ref{fig_SBF}(b) and (h)).
Then the fluctuation images are obtained by subtracting median-filtered galaxy images from the original galaxy images and normalizing them by the square root of the filtered images.

Using a median-filtered image as a substitute for galaxy model images is not conventional, but we could minimize the differences between the two methods by using an appropriate filter size.  
We tested this using galaxies that are well described by a single S\'ersic model, such as BK5N (Figure \ref{fig_SBF}a).
In this case we derived a smooth galaxy model image using \texttt{GALFIT} \citep{pen02} as well as a median-filtered galaxy image. Then we obtained  azimuthally averaged power spectra of the fluctuation images, which are shown in Figure \ref{fig_SBF}(e): black dots for the median-filtered galaxy image and gray thick lines for the \texttt{GALFIT} modelled galaxy image. 
Both agree very well except for the largest scale ($k<0.03$). 
We tried multiple filter sizes to derive 
power spectra, 
and found the best match with those 
based on \texttt{GALFIT} models when the filter size = $10\times$ seeing size. 
Using too small a filter size removes some of PSF-scale fluctuations and using too large a filter size results in underestimating the galaxy light.
Therefore, we use median filtered images as smooth galaxy models, assuming that they do not show any significant difference in their power spectra compared with  standard methods. All the model images and power spectra in Figure \ref{fig_SBF} are derived with filter size being ten times the seeing size.

\subsection{Masking Contaminating Sources}\label{sec_sbf_mask}

Essentially, masks are applied in order to exclude regions that contaminate true galaxy fluctuation signals or result in increasing measurement errors.
There are two kinds of masks needed: 
(1) annular masks for area selection and 
(2) individual source masks for contaminating sources.

\subsubsection{Annular Mask for Area Selection}

We choose an optimal area for SBF calculation of each field to minimize measurement errors.
If one uses too large an area including the low surface brightness region, $P_0$ fluctuations from contaminating sources as well as from true SBF signals are amplified.
The background subtraction procedure can mitigate this effect since the fluctuation power from background sources (e.g. background galaxies) increases as well.
However, errors in background subtraction unnecessarily increase in this case.
Also, using too small an area is not recommended because it results in poor fits to the power spectrum due to the small number of pixels. 
So, we measured the SBF using various annular masks and selected an area based on the galaxy's surface brightness and size.
Solid line ellipses in Figure \ref{fig_sample} are annular mask boundaries of inner fields and dotted line rings are those of outer fields.
The 
color and mean surface brightness of the selected areas for SBF measurements of the sample galaxies are 
listed in Table \ref{tab_calibration}.

\subsubsection{Individual Source Mask}

There are two kinds of contaminating sources that should not be included in calculating the SBF: (1) background objects such as faint galaxies and foreground Milky Way stars, and (2) young stellar populations of the galaxy.
Contribution from background objects can be successfully taken into account in two steps as described in \citet{car19}. 
First, objects brighter than a masking threshold are masked when calculating galaxy fluctuation power $P_0$. Second, contributions from objects fainter than the threshold are statistically estimated and subtracted by measuring the background fluctuations in flat fields well away from the galaxy, within a radius of 0.2 degrees
(PSF-scale components in Figure \ref{fig_SBF}(f) and (l)), which are masked using the same threshold. When using a brighter masking threshold, one would obtain a larger galaxy fluctuation and a larger background fluctuation, so the difference should remain approximately constant. Therefore, the choice of masking threshold does not yield any significant systematic uncertainties in the SBF measurement of red galaxies unless one uses a faint magnitude threshold that masks the galaxy's asymptotic giant branch (AGB) and red giant branch (RGB) stars, which produce the desired SBF signal.

In contrast, young stellar populations are trickier to handle and the choice of masking thresholds may lead to systematic uncertainties. Young massive blue stars are small in number, but they are very bright and are clumped in star forming regions, contributing to large fluctuations. Therefore, they result in significant SBF and stochastic effects and it is important to avoid including star forming regions in SBF calculation. 

\begin{figure*}[hbt!]
\centering
\includegraphics[scale=0.9]{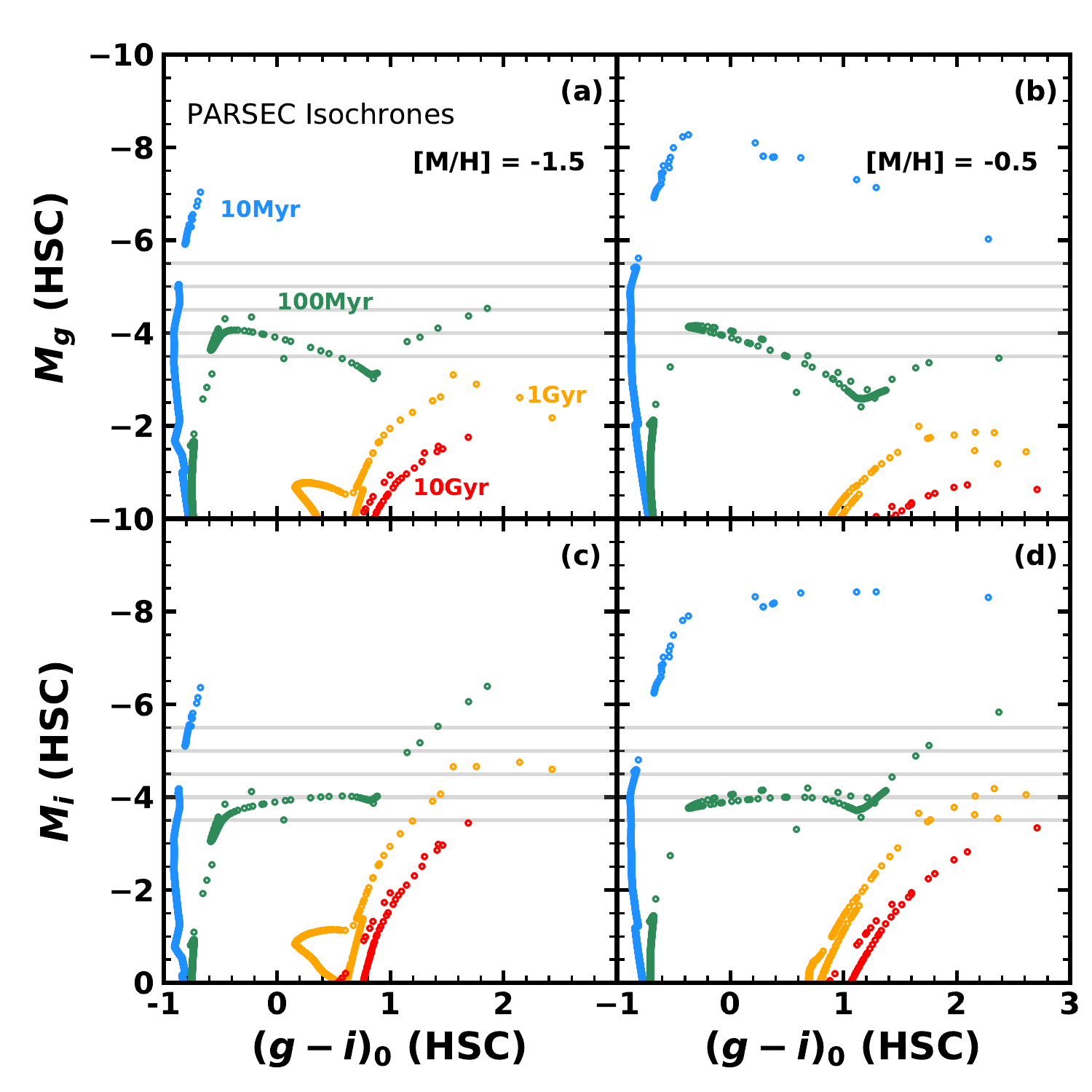} 
\caption{
Color-magnitude diagrams of simulated populations with a total mass of $10^5{\rm M_{\odot}}$ from \texttt{PARSEC} evolutionary tracks. 
Upper panels (a and b) represent $M_g$ versus $(g-i)_0$ and lower panels (c and d) represent $M_i$ versus $(g-i)_0$. Left panels (a and c) and right panels (b and d) display metal-poor ([M/H] = --1.5) and metal-rich ([M/H] = --0.5) populations, respectively.
Colors indicate stellar isochrones for four ages (10 Myr, 100 Myr, 1 Gyr, and 10 Gyr). 
Gray horizontal lines indicate the five masking thresholds ($M_{g,{\rm thres}}$) used in this study.
} 
\label{fig_threshold}
\end{figure*}

To show the effect of young stellar populations, we plot, in  Figure \ref{fig_threshold}, color-magnitude diagrams of simulated simple stellar populations with a total mass of $10^5{\rm M_{\odot}}$ from \texttt{PARSEC} evolutionary tracks \citep{bre12}\footnote{http://stev.oapd.inaf.it/cmd}, for ages 10 Myr, 100 Myr, 1 Gyr, and 10 Gyr.
\texttt{PARSEC} tracks are a revised version of the former Padova tracks, computed for a scaled-solar composition and following the helium abundance $Y = 0.2485 + 1.78 Z$. \texttt{PARSEC} computes stellar evolution from the pre main-sequence to the onset of the first thermal pulsation. The evolution of TP-AGB stars is traced with \texttt{COLIBRI} codes \citep{mar13, mar17}, from the first thermal pulsation to the total loss of the envelope.
Circles in the left panels (a and c) and in the right panels (b and d)
represent metal-poor ([M/H] = --1.5) and metal-rich populations ([M/H] = --0.5), respectively. 
Stellar populations that are bright in $i-$band are the main contributors to the $i-$band SBF.
If a galaxy lacks young stellar populations and consists of intermediate-age and old stellar populations, RGB and AGB stars dominate its SBF. 
However, if young stellar populations are present, massive main sequence stars, core helium burning stars, and supergiant stars would contribute to larger fluctuations.
A majority of the young stellar populations
have bluer colors than older stellar populations so they are more prominent in $g$-band magnitudes than in $i$-band magnitudes.

We check for systematic effects with five masking thresholds:
$M_g = -3.5, -4.0, -4.5, -5.0,$ and $-5.5$ mag, indicated as gray horizontal lines in Figure \ref{fig_threshold}. We use \texttt{SExtractor} \citep{ber96} to detect contaminating objects. 
All the objects brighter than a threshold are masked using segmentation maps 
generated by the \texttt{SExtractor}, which indicate the pixels belonging to each detected source. 
Masking objects in the range $-5.5 \leq M_g \leq -3.5$ would mask unresolved young star clusters, bright blue main-sequence stars, core helium burning stars, and supergiant stars, while leaving intermediate-age AGB stars and old RGB stars unmasked. 
However, absolute $i-$band magnitudes of intermediate-age AGB stars lie in the range $M_i \lesssim -4$, and SBFs are expected to vary significantly according to the choice of masking thresholds in $i-$band magnitudes. 
Therefore, using $g-$band images to detect contaminating sources is favored over using $i-$band images.
Red objects that are fainter than the $g-$band threshold but do not look like fluctuation within the galaxy are additionally masked by visual inspection. 

With the chosen masking threshold, objects in nearby background fields are masked as well. 
The background fluctuation power ($P_{0, back}$) is the PSF-scaled power ($P_0$ in Equation \ref{ps}) of ${\rm (Background)/ \sqrt{Model}}$.
We calculate $P_{0, back}$ in 100 random nearby background areas, and adopt its median value.
Then we obtain background-subtracted SBF power by subtracting the median $P_{0, back}$ from $P_0$.
We use $1\sigma$ range of $P_{0, back}$ as a background subtraction error.

Small contours in Figure \ref{fig_SBF}(a) and (g) represent masked sources. The dark blue, light blue, cyan, light green, and orange contours indicate sources masked using thresholds of $M_g = -3.5, -4.0, -4.5, -5.0,$ and $-5.5$ mag. 
While BK5N shows only a few bright blue sources,
in the central elliptical area used for calculating the SBF, DDO 6, has many such sources with various magnitudes.
The effects of using different masking thresholds are discussed in Section \ref{sec_calibration_threshold}.

\subsection{Power Spectrum Fitting}\label{sec_sbf_fit}

We fit the power spectra (right panels of Figure \ref{fig_SBF}) to the linear function (Equation \ref{ps}) by the least-squares method. 
Dashed red lines represent the PSF-scale fluctuation component $P_0E(k)$ and dotted lines represent the white noise component ($P_1$).
The lowest wavenumbers and the highest wavenumbers (gray shaded regions in right panels in Fig \ref{fig_SBF}) are excluded from the fitting range.
This is because the lowest wavenumbers are affected by large scale subtraction errors due to imperfect galaxy models and the highest wavenumbers are affected by pixel interpolations introduced when stacking images.
The minimum wavenumber is chosen typically in the range [0.03, 0.08] pix$^{-1}$ and the maximum wavenumber is chosen in the range [0.3, 0.4] pix$^{-1}$.
We iterate the fitting procedure to find the best ranges. 
The uncertainty introduced by variable fitting range is determined as one-sigma range of results from multiple fits. 
Measurement errors of SBF correspond to the quadratic sum of background subtraction error and power spectrum fitting error.

\begin{figure}[bt!]
\centering
\includegraphics[scale=0.58]{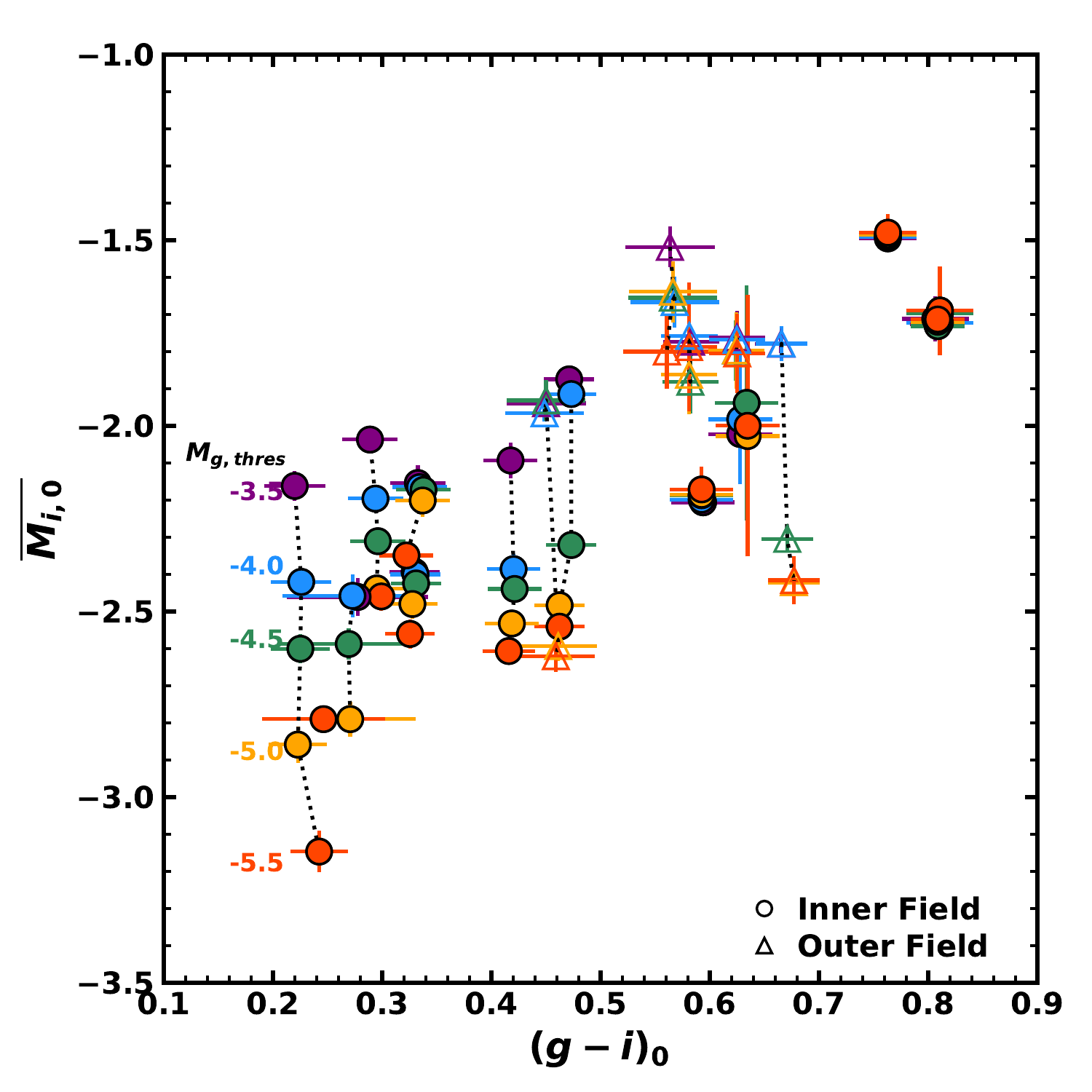} 
\caption{
SBF absolute magnitude versus integrated color for the calibration sample galaxies.
Symbol colors indicate masking thresholds, of which red, orange, green, blue, purple colors represent $M_{g,{\rm thres}} = -5.5, -5.0, -4.5, -4.0,$ and $-3.5$ mag respectively. Circles and triangles indicate inner and outer regions of the galaxies, respectively.}
\label{fig_calibration_all}
\end{figure}

\begin{figure*}[hbt!]
\centering
\includegraphics[scale=0.6]{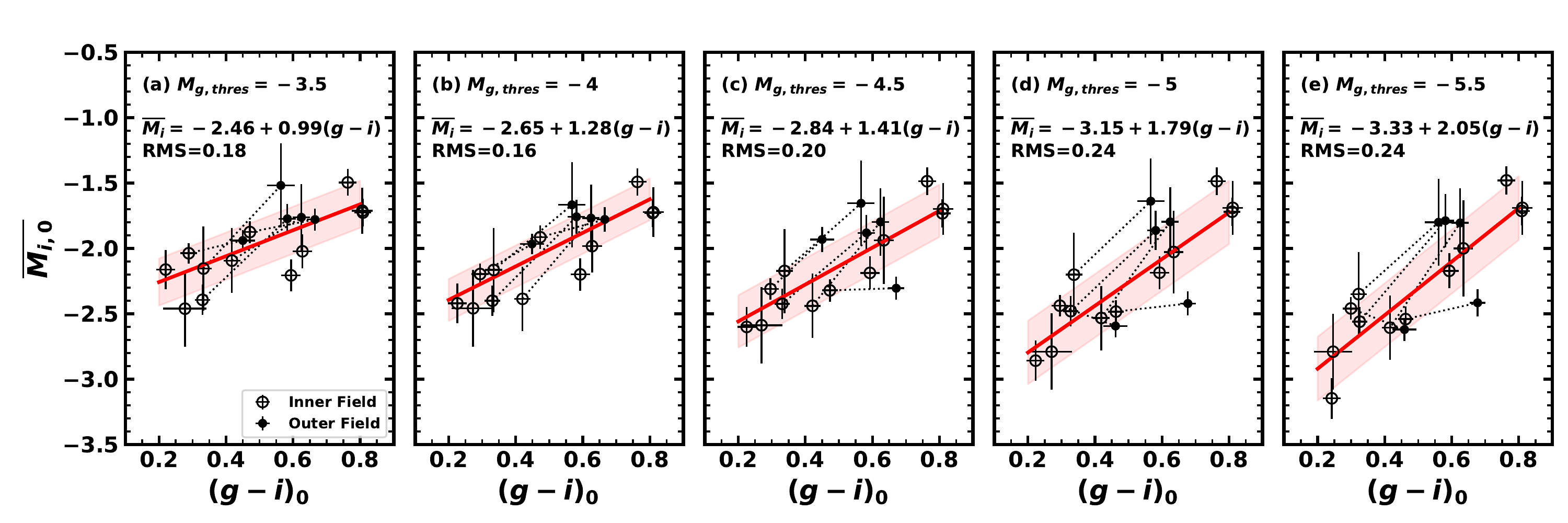} 
\caption{SBF and integrated color relations of the calibration sample, with $M_{g,{\rm thres}}=-3.5, -4.0, -4.5, -5.0,$ and $-5.5$ mag. Open circles and filled circles indicate inner and outer regions of the galaxies, respectively. Linear fits are shown as red lines. 
}
\label{fig_calibration}
\end{figure*}


\section{SBF Calibration}\label{sec_calibration}

We derive the SBF apparent magnitudes of 17 fields in 12 dwarf galaxies 
using the method described in the previous section. Using the TRGB distances to the galaxies, we obtain the SBF absolute magnitudes, $\overline{M_i}$. The results are listed in Table \ref{tab_calibration} . The listed errors in $\overline{M_i}$ include SBF measurement errors. 
In this section, we derive SBF and integrated color relations for the HSC $i-$band. 
But first, we check how the choice of masking thresholds $M_{g,{\rm thres}}$ affects SBF magnitudes. 

\subsection{Dependence on Masking Thresholds}\label{sec_calibration_threshold}

Figure \ref{fig_calibration_all} shows SBF and integrated color relations for our 17 fields. 
Circular symbols and triangle symbols indicate inner and outer regions of the galaxies, respectively. Symbol colors represent $M_{g,{\rm thres}}$, of which red, orange, green, blue, and purple colors show --5.5, --5.0, --4.5, --4.0, and --3.5 mag. 
In general, using fainter $M_{g,{\rm thres}}$ results in fainter fluctuations. While SBF magnitudes of most of the fields in the color range $(g-i)_0 \gtrsim 0.6$ do not vary significantly according to $M_{g,{\rm thres}}$, bluer fields show a larger variation. This is because bluer galaxies have more sources in the magnitude range between $-5.5 < M_g < -3.5$, mostly young stellar populations. 

\subsection{Fitting SBF Magnitude--Color Relations}

\begin{deluxetable}{ccccc}[bt!]
\tabletypesize{\footnotesize}
\tablecaption{Fluctuation - Integrated Color Relation Fitting Results   \label{tab_fit} }
\tablewidth{0pt}
\tablehead{&\multicolumn{3}{c}{
$y_i=\alpha+\beta\times x_i + \epsilon_i$,   $\epsilon_i \sim N(0,\sigma_{int}^2)$
} & \\
\colhead{$M_{g,{\rm thres}}$}  & \colhead{$\alpha$} & \colhead{$\beta$} & \colhead{$\sigma_{int}$} & \colhead{rms} 
}
\startdata
--3.5 &  $-2.46\pm0.14$  &  $1.01\pm0.24$  &  $0.11\pm0.05$  & 0.18 \\ 
--4.0 &  $-2.65\pm0.13$  &  $1.28\pm0.24$  &  $0.09\pm0.05$  & 0.16 \\ 
--4.5 &  $-2.84\pm0.18$  &  $1.41\pm0.32$  &  $0.18\pm0.06$  & 0.20 \\ 
--5.0 &  $-3.15\pm0.20$  &  $1.79\pm0.36$  &  $0.21\pm0.06$  & 0.24 \\ 
--5.5 &  $-3.33\pm0.20$  &  $2.08\pm0.38$  &  $0.21\pm0.07$  & 0.24 \\ 
\enddata
\end{deluxetable}

In Figure \ref{fig_calibration} 
we plot SBF absolute magnitude versus integrated color of the dwarf galaxies for five different masking thresholds.
The open circles and filled circles indicate measurements of inner and outer regions, respectively. Dotted lines link the measurements of the same galaxy. In general, the fluctuation magnitudes in the inner region are brighter than in the outer region. This radial gradient SBF gradient is likely to be resulting from an age effect, as suggested by \citet{jen15}. Young stellar populations in the central region enhance the SBF signals.

We model the fluctuation magnitude and color relation by a linear function with a gaussian intrinsic scatter ($\sigma_{int}$): 
$y_i=\alpha+\beta\times x_i + \epsilon_i$. 
$\epsilon_i$ is assumed to follow a normal distribution with a mean 0 and a standard deviation $\sigma_{int}$.
Then we use \texttt{emcee} \citep{for13}, a python implementation of Monte-Carlo Markov Chain ensemble sampler for sampling posterior distributions.
We 
consider errors in both axes, $x=(g-i)_0$ and $y=\overline{M_i}$. 
$(g-i)_0$ errors are the quadratic sum of Poisson errors within the area used for SBF measurement, sky fluctuations, and the photometric calibration error 0.017 mag \citep{aih19} per filter.
$\overline{M_i}$ errors used for the fitting are the quadratic sum of SBF measurement errors and the TRGB distance errors. 
Table \ref{tab_fit} lists fitting results.
Red lines in Figure \ref{fig_calibration} indicate linear fits and shaded red regions show corresponding rms scatter.

In Figure \ref{fig_calibration}, the slope of the relation decreases and the y-intercept increases as fainter masking thresholds are used. 
The rms scatter is the smallest (0.16 mag) when using $M_{g,{\rm thres}} = -4.0$ and it increases with brighter or fainter masking thresholds.
Therefore, the slope and rms scatter of the relation vary depending on how many young stars are masked. If more such sources are masked, the slope and rms scatter decrease.
For the case using $M_{g,{\rm thres}} = -4.0$, the intrinsic scatter is found to be
$\sigma_{int}=0.09\pm0.05$.
The value of $\sigma_{int}$ depends on the values of input errors of $(g-i)_0$ and $\overline{M_i}$. The $(g-i)_0$ errors used in this study are likely lower bounds, since we do not quantify errors in global sky subtraction by \texttt{hscpipe}. Thus, the value of $\sigma_{int}$ is likely overestimated.

Considering the minimum rms scatter, we choose output parameters 
for $M_{g,{\rm thres}} = -4.0$ as our final calibration:
\begin{equation}\label{calibration}
\overline{M_i} = (-2.65 \pm 0.13) + (1.28 \pm 0.24) \times (g-i)_0.
\end{equation}

\section{Comparison with Previous Calibrations and Stellar Population Models}\label{sec_discussion}

\begin{figure*}[hbt!]
\centering
\includegraphics[scale=0.6]{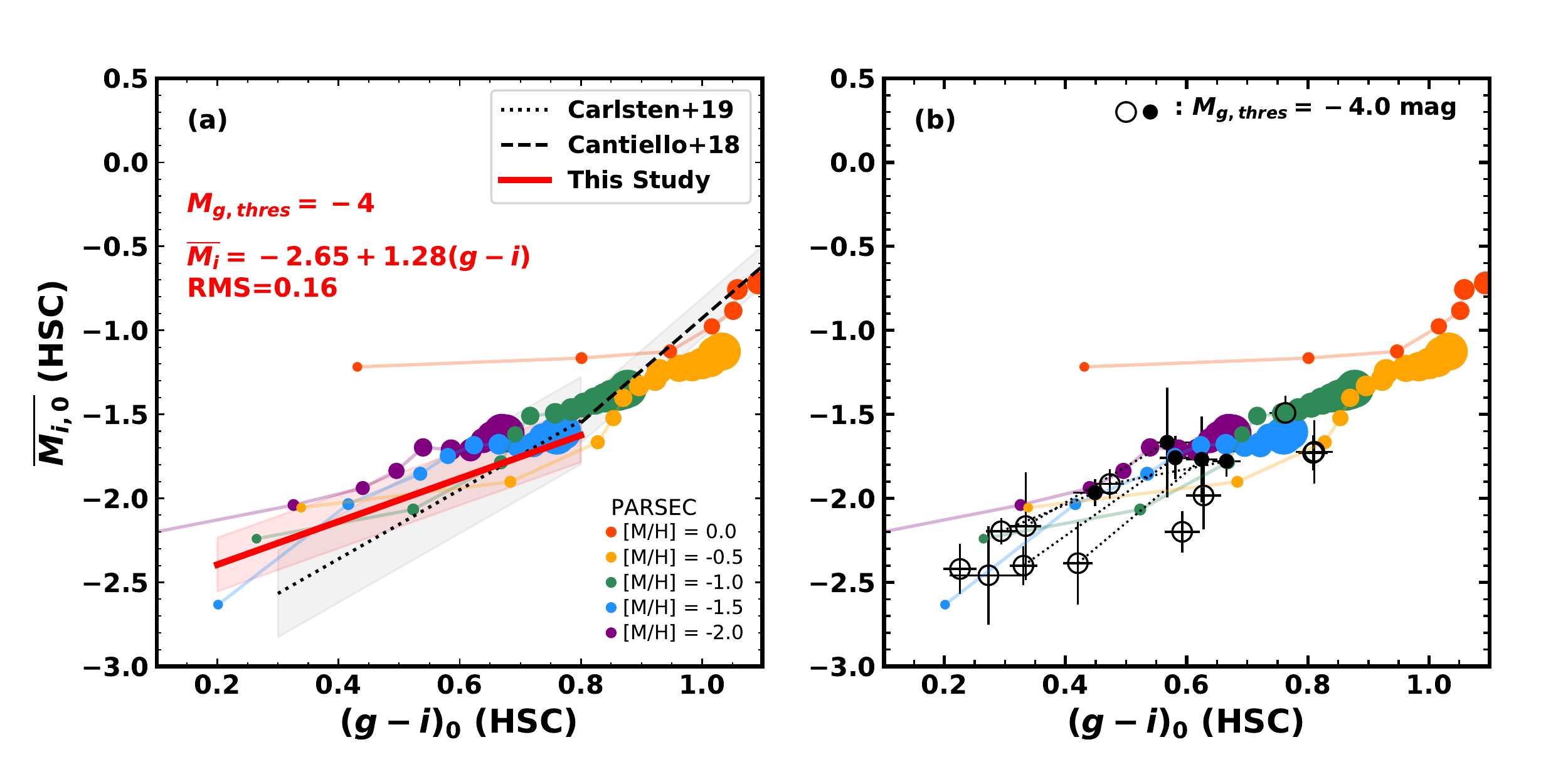} 
\caption{(a) Comparison of our SBF calibration for $M_{g,{\rm thres}}=-4.0$ mag (red line) with recent calibrations in the literature \citep[as black lines][]{car19, can18} and \texttt{PARSEC} models for simple stellar populations. 
The orange, yellow, green, blue, and purple circles 
indicate five different metallicities, [M/H] = $0.0, -0.5, -1.0, -1.5$, and $-2.0$. 
Circle sizes indicate relatively ages in the range [1, 14] Gyr, in a linear spacing, with larger circles being older. 
(b) Same as the left but with our measured SBFs of the calibration sample as open black (inner fields) and filled black (outer fields) symbols with errorbars. } 
\label{fig_calibration2}
\end{figure*}

Figure \ref{fig_calibration2} 
compares our SBF calibrations with two recent ground-based empirical SBF calibrations \citep{can18, car19} and \texttt{PARSEC} stellar population models.
The SBF absolute magnitudes of stellar population models are computed as luminosity-weighted mean luminosities by integrating weighted stellar isochrones.
Our calibration for $M_{g,{\rm thres}}=-4.0$ mag is shown as a red line 
in Figure \ref{fig_calibration2}(a).
Our data are plotted with black circles in Figure \ref{fig_calibration2}(b). 
\citet{can18}, as part of The Next Generation Virgo Cluster Survey (NGVS), used bright red galaxies with $0.8 < g-i < 1.1$ in the Virgo cluster observed with the CFHT to derive an $i$-band calibration 
(black dashed line). 
\citet{car19} derived 
an $i$-band calibration in a blue regime ($0.3 < g-i < 0.8$) using 32 LV dwarf galaxies (28 of which are based on CFHT archival data and 6 of which are based on HST data in \citet{coh18}). They showed that the slope of the SBF calibration is shallower in a bluer, metal-poor regime (black dotted line).
Since these two calibrations are obtained in the CFHT magnitude system, we transformed them to HSC magnitudes by
comparing CFHT and HSC $g, i$ magnitudes from \texttt{PARSEC} isochrones. While the $i-$band magnitudes of the two systems are almost identical in the color range $0.0 < (g-i) < 1.5$, their $g-$band magnitudes differ slightly, depending on color: 
$g_{\rm HSC} = g_{\rm CFHT} + 0.045 (g-i)_{\rm CFHT} + 0.006$ (rms=0.002 mag).  
Using these relations, we obtain  $\overline{M_{i,{\rm HSC}}} = -3.18 + 2.06\times(g-i)_{0,{\rm HSC}}$ with rms = 0.26 mag for \citet{car19} and $\overline{M_{i,{\rm HSC}}} = -4.04 + 3.11 (g-i)_{0,{\rm HSC}}$ with rms = 0.12 mag for \citet{can18}.

Compared to the slope of $\beta = 2.06 \pm 0.39$ derived by \citet{car19} 
for the color range similar to that in this study, 
our derived slope $\beta = 1.28 \pm 0.24$ is slightly shallower.
Also, the rms scatter of our calibration (0.16 mag) is much smaller than that of \citet{car19}, 0.26 mag. 
Among our five calibrations with different masking thresholds, the slopes obtained using $M_{g,{\rm thres}} = -5.0$ and $-5.5$ mag ($\beta = 1.79\pm0.36$ and $2.08\pm0.38$) are similar to that of \citet{car19}. 
In addition, the rms scatter of the two cases, 0.24 mag, is similar to that of \citet{car19}, 0.26 mag.
Therefore, we conjecture that the average masking done by \citet{car19} is similar to our case of $M_{g,{\rm thres}} = -5.0$ or $-5.5$ mag.

As suggested by previous studies \citep{mie06, bla09, jen15, car19}, the color dependence of SBF absolute magnitude on colors is smaller in the blue regime. Comparing our $\beta$ with the slope of \citet{can18} ($\beta = 3.11 \pm 0.42$) which presented an $i-$band SBF calibration with $g-i$ colors for the red regime, our slope is significantly shallower.
At the color between the blue and red regime, $(g-i)_0=0.8$,
our calibration is only $\sim$0.07 mag brighter than that of \citet{can18}, so both calibrations are consistent within the error range.

In Figure \ref{fig_calibration2}, the colored circles display theoretical models calculated from \texttt{PARSEC} simple stellar population models, with ages [1, 14] Gyr and metallicities ([M/H]) [--2.0, 0.0]. 
Symbol colors vary depending on metallicity, and symbol sizes vary depending on relative age.
The red regime ($(g-i)_0 \gtrsim 0.8$ mag) corresponds to metal-rich populations and the models for old ages and high metallicities show a fair agreement with \citet{can18}.
In the blue regime ($(g-i)_0 \lesssim 0.8$ mag), where both old metal-poor populations and intermediate-age metal-rich populations may be present, the scatter of $\overline{M_i}$ between models is larger.
Generally, the empirical $\overline{M_i}$ for the blue regime is 
consistent with simple stellar models for low metallicity. 
The dwarf galaxies at $0.6\lesssim(g-i)_0\lesssim0.8$ are consistent with models for old ages, and those at $0.2\lesssim(g-i)_0\lesssim0.6$ overlap with models for younger ages (a few Gyrs).
Despite the potential complicated effects that might result from unmasked young stellar populations in late-type galaxies (see \citet{gre21} for a detailed study on effects of young stellar populations), our empirical calibration shows a reasonable agreement with predictions from simple stellar population models. 

\section{Summary}\label{sec_summary}

We have derived a new empirical $i-$band SBF calibration valid in a blue regime ($0.2 \lesssim (g-i)_0 \lesssim 0.8$) in the HSC magnitude system, using 12 dwarf galaxies ($D_{\rm TRGB} < 10$ Mpc) in the SMOKA Science Archive.
The sample is composed of various morphological types of dwarf galaxies, from star-forming irregular galaxies to dwarf spheroidal galaxies.

Measuring SBFs of dwarf galaxies with young stellar populations is tricky and requires modifications to standard methods.
We described the procedures of SBF analysis, applicable for blue, star-forming galaxies. In making smooth galaxy models, we used median-filtered images with the filter size of ten times the seeing size instead of fitting a S\'ersic profile. In addition, we masked young stellar populations brighter than a $g-$band masking threshold, $M_{g,{\rm thres}}$. 

We tried multiple masking thresholds, $M_{g,{\rm thres}} = -3.5, -4.0, -4.5, -5.0$, and $-5.5$ mag and derived SBF calibrations for each case, as summarized in Table \ref{tab_fit}. 
Using a fainter $M_{g,{\rm thres}}$, the fluctuations of galaxies in the blue regime systematically decrease and the slope of the SBF calibration gets shallower. In the case of $M_{g,{\rm thres}} = -4.0$ mag, the rms scatter of the linear fit is the smallest, 0.16 mag (Equation \ref{calibration}).
This scatter is much smaller than those in the previous studies for dwarf galaxies in the blue regime.
Our result shows a reasonable agreement with \texttt{PARSEC} models for metal-poor, intermediate and old simple stellar populations.

Recently \citet{kim21} estimated the SBF distances to the dwarf galaxy candidates around NGC 4437 from the HSC data, applying the calibration in this study, and they could identify successfully the members of the NGC 4437 group.
The calibration in this study will be very useful for estimating distances to dwarf galaxies in the LV found in the various deep and wide surveys.

\acknowledgments
We thank the anonymous referee for providing helpful comments. 
This work  was supported by the National Research Foundation grant funded by the Korean Government (NRF-2019R1A2C2084019).
We thank Brian S. Cho for his help in improving the English in this manuscript.


\clearpage


\begin{thebibliography}{}


\bibitem[Aihara et al.(2019)]{aih19} Aihara, H., AlSayyad, Y., Ando, M., et al.\ 2019, \pasj, 71, 114

\bibitem[Anand et al.(2021)]{ana21} Anand, G.~S., Rizzi, L., Tully, R.~B., et al.\ 2021, \aj, 162, 80. doi:10.3847/1538-3881/ac0440

\bibitem[Bertin \& Arnouts(1996)]{ber96} Bertin, E. \& Arnouts, S.\ 1996, \aaps, 117, 393. doi:10.1051/aas:1996164

\bibitem[Blakeslee et al.(2001)]{bla01} Blakeslee, J.~P., Vazdekis, A., \& Ajhar, E.~A.\ 2001, \mnras, 320, 193. doi:10.1046/j.1365-8711.2001.03937.x

\bibitem[Blakeslee et al.(2009)]{bla09} Blakeslee, J.~P., Jord{\'a}n, A., Mei, S., et al.\ 2009, \apj, 694, 556. doi:10.1088/0004-637X/694/1/556

\bibitem[Blakeslee et al.(2010)]{bla10} Blakeslee, J.~P., Cantiello, M., Mei, S., et al.\ 2010, \apj, 724, 657. doi:10.1088/0004-637X/724/1/657

\bibitem[Blakeslee et al.(2021)]{bla21} Blakeslee, J.~P., Jensen, J.~B., Ma, C.-P., et al.\ 2021, \apj, 911, 65. doi:10.3847/1538-4357/abe86a

\bibitem[Bosch et al.(2018)]{bos18} Bosch, J., Armstrong, R., Bickerton, S., et al.\ 2018, \pasj, 70, S5. doi:10.1093/pasj/psx080

\bibitem[Bressan et al.(2012)]{bre12} Bressan, A., Marigo, P., Girardi, L., et al.\ 2012, \mnras, 427, 127. doi:10.1111/j.1365-2966.2012.21948.x

\bibitem[Bullock \& Boylan-Kolchin(2017)]{bul17} Bullock, J.~S. \& Boylan-Kolchin, M.\ 2017, \araa, 55, 343. doi:10.1146/annurev-astro-091916-055313

\bibitem[Cantiello et al.(2005)]{can05} Cantiello, M., Blakeslee, J.~P., Raimondo, G., et al.\ 2005, \apj, 634, 239. doi:10.1086/491694

\bibitem[Cantiello et al.(2018)]{can18} Cantiello, M., Blakeslee, J.~P., Ferrarese, L., et al.\ 2018, \apj, 856, 126

\bibitem[Carlin et al.(2016)]{carlin16} Carlin, J.~L., Sand, D.~J., Price, P., et al.\ 2016, \apjl, 828, L5. doi:10.3847/2041-8205/828/1/L5

\bibitem[Carlin et al.(2021)]{carlin21} Carlin, J.~L., Mutlu-Pakdil, B., Crnojevi{\'c}, D., et al.\ 2021, \apj, 909, 211. doi:10.3847/1538-4357/abe040


\bibitem[Carlsten et al.(2019a)]{car19a} Carlsten, S.~G., Beaton, R.~L., Greco, J.~P., et al.\ 2019, \apjl, 878, L16. doi:10.3847/2041-8213/ab24d2

\bibitem[Carlsten et al.(2019b)]{car19} Carlsten, S.~G., Beaton, R.~L., Greco, J.~P., et al.\ 2019, \apj, 879, 13


\bibitem[Carlsten et al.(2020a)]{car20a} Carlsten, S.~G., Greco, J.~P., Beaton, R.~L., et al.\ 2020, \apj, 891, 144. doi:10.3847/1538-4357/ab7758 

\bibitem[Carlsten et al.(2021)]{car21} Carlsten, S.~G., Greene, J.~E., Peter, A.~H.~G., et al.\ 2021, \apj, 908, 109. doi:10.3847/1538-4357/abd039


\bibitem[Cohen et al.(2018)]{coh18} Cohen, Y., van Dokkum, P., Danieli, S., et al.\ 2018, \apj, 868, 96. doi:10.3847/1538-4357/aae7c8

\bibitem[Davis et al.(2021)]{dav21} Davis, A.~B., Nierenberg, A.~M., Peter, A.~H.~G., et al.\ 2021, \mnras, 500, 3854. doi:10.1093/mnras/staa3246

\bibitem[Foreman-Mackey et al.(2013)]{for13} Foreman-Mackey, D., Hogg, D.~W., Lang, D., et al.\ 2013, \pasp, 125, 306

\bibitem[Greco et al.(2021)]{gre21} Greco, J.~P., van Dokkum, P., Danieli, S., et al.\ 2021, \apj, 908, 24. doi:10.3847/1538-4357/abd030

\bibitem[Ivezi{\'c} et al.(2019)]{ive19} Ivezi{\'c}, {\v{Z}}., Kahn, S.~M., Tyson, J.~A., et al.\ 2019, \apj, 873, 111

\bibitem[Jacobs et al.(2009)]{jac09} Jacobs, B.~A., Rizzi, L., Tully, R.~B., et al.\ 2009, \aj, 138, 332. doi:10.1088/0004-6256/138/2/332

\bibitem[Jang et al.(2021)]{jan21} Jang, I.~S., Hoyt, T.~J., Beaton, R.~L., et al.\ 2021, \apj, 906, 125. doi:10.3847/1538-4357/abc8e9

\bibitem[Jensen et al.(1998)]{jen98} Jensen, J.~B., Tonry, J.~L., \& Luppino, G.~A.\ 1998, \apj, 505, 111. doi:10.1086/306163

\bibitem[Jensen et al.(2003)]{jen03} Jensen, J.~B., Tonry, J.~L., Barris, B.~J., et al.\ 2003, \apj, 583, 712. doi:10.1086/345430

\bibitem[Jensen et al.(2015)]{jen15} Jensen, J.~B., Blakeslee, J.~P., Gibson, Z., et al.\ 2015, \apj, 808, 91. doi:10.1088/0004-637X/808/1/91

\bibitem[Jerjen et al.(1998)]{jer98} Jerjen, H., Freeman, K.~C., \& Binggeli, B.\ 1998, \aj, 116, 2873

\bibitem[Jerjen et al.(2000)]{jer00} Jerjen, H., Freeman, K.~C., \& Binggeli, B.\ 2000, \aj, 119, 166. doi:10.1086/301188

\bibitem[Jerjen et al.(2001)]{jer01} Jerjen, H., Rekola, R., Takalo, L., et al.\ 2001, \aap, 380, 90. doi:10.1051/0004-6361:20011408


\bibitem[Karachentsev et al.(2013)]{kar13b} Karachentsev, I.~D., Makarov, D.~I., \& Kaisina, E.~I.\ 2013, \aj, 145, 101. doi:10.1088/0004-6256/145/4/101

\bibitem[Kim et al.(2021)]{kim21} Kim, Y.~J., Kang, J., Lee, M.~G., \& Jang, I.~S.\ 2021, \apj, submitted

\bibitem[Klypin et al.(1999)]{kly99} Klypin, A., Kravtsov, A.~V., Valenzuela, O., et al.\ 1999, \apj, 522, 82. doi:10.1086/307643

\bibitem[Lee et al.(1993)]{lee93} Lee, M.~G., Freedman, W.~L., \& Madore, B.~F.\ 1993, \apj, 417, 553 

\bibitem[Liu et al.(2000)]{liu00} Liu, M.~C., Charlot, S., \& Graham, J.~R.\ 2000, \apj, 543, 644. doi:10.1086/317147

\bibitem[Marigo et al.(2013)]{mar13} Marigo, P., Bressan, A., Nanni, A., et al.\ 2013, \mnras, 434, 488. doi:10.1093/mnras/stt1034

\bibitem[Marigo et al.(2017)]{mar17} Marigo, P., Girardi, L., Bressan, A., et al.\ 2017, \apj, 835, 77. doi:10.3847/1538-4357/835/1/77

\bibitem[Mei et al.(2005)]{mei05} Mei, S., Blakeslee, J.~P., Tonry, J.~L., et al.\ 2005, \apjs, 156, 113

\bibitem[Mieske et al.(2006)]{mie06} Mieske, S., Hilker, M., \& Infante, L.\ 2006, \aap, 458, 1013. doi:10.1051/0004-6361:20054685

\bibitem[Moore et al.(1999)]{moo99} Moore, B., Ghigna, S., Governato, F., et al.\ 1999, \apjl, 524, L19. doi:10.1086/312287

\bibitem[M{\"u}ller \& Jerjen(2020)]{mul20} M{\"u}ller, O. \& Jerjen, H.\ 2020, \aap, 644, A91. doi:10.1051/0004-6361/202038862

\bibitem[Peng et al.(2002)]{pen02} Peng, C.~Y., Ho, L.~C., Impey, C.~D., et al.\ 2002, \aj, 124, 266. doi:10.1086/340952

\bibitem[Schlafly \& Finkbeiner(2011)]{sch11} Schlafly, E.~F., \& Finkbeiner, D.~P.\ 2011, \apj, 737, 103 

\bibitem[Schlegel et al.(1998)]{sch98} Schlegel, D.~J., Finkbeiner, D.~P., \& Davis, M.\ 1998, \apj, 500, 525

\bibitem[Smercina et al.(2018)]{sme18} Smercina, A., Bell, E.~F., Price, P.~A., et al.\ 2018, \apj, 863, 152. doi:10.3847/1538-4357/aad2d6

\bibitem[Tonry \& Schneider(1988)]{ton88} Tonry, J., \& Schneider, D.~P.\ 1988, \aj, 96, 807

\bibitem[Tonry et al.(1990)]{ton90} Tonry, J.~L., Ajhar, E.~A., \& Luppino, G.~A.\ 1990, \aj, 100, 1416. doi:10.1086/115606

\bibitem[Tully et al.(2009)]{tul09} Tully, R.~B., Rizzi, L., Shaya, E.~J., et al.\ 2009, \aj, 138, 323. doi:10.1088/0004-6256/138/2/323

\bibitem[van Dokkum et al.(2018)]{van18} van Dokkum, P., Danieli, S., Cohen, Y., et al.\ 2018, \nat, 555, 629. doi:10.1038/nature25767


\end{thebibliography}
\end{document}